\documentclass{article}
\usepackage[utf8]{inputenc}
\usepackage{graphicx}
\graphicspath{ {./} }
\usepackage{authblk}
\usepackage{textcomp}
\usepackage{cite} 
\usepackage[numbers, sort]{natbib}
\usepackage{epstopdf}
\usepackage{bm}% bold math
\usepackage[T1]{fontenc}
\usepackage{booktabs, array, mathptmx, float, tabularx, booktabs, lipsum, amsmath,multirow}
\usepackage{siunitx, xcolor}
\usepackage[version=4]{mhchem}
\usepackage{hyperref}
\hypersetup{colorlinks=true,linkcolor=blue,anchorcolor=blue,urlcolor=blue,citecolor=blue}

\title{Effect of atomic anti-site disorder on the AMR in FeCo alloys}

\author[1]{Mingsong Zhang}
\author[1]{Bin Peng}
\author[1]{Wanli Zhang}
\author[1]{Wenxu Zhang \thanks{Corresponding Author: xwzhang@uestc.edu.cn}  }

\affil[1]{School of Integrated Circuits Science and Engineering, University of Electronic Science and Technology of China, 610081, Chengdu, P.R. China}

\date{}

\begin{document}
	
\maketitle
	
\begin{abstract}
In order to understand the anti-site disorder effect on the anisotropic magnetoresistance (AMR) effect in alloys, $\rm{Fe}_{50}Co_{50}$ alloys were studied in this work using the fully relativistic spin-polarized screened (KKR) method. The anti-site effect was modeled by interchanging Fe and Co atoms and treated by the coherent potential approximation (CPA). We find that the anti-site disorder broadens the spectral function and decreases the conductivity. Our work emphasizes that the absolute variations of resistivity under magnetic moment rotation are less affected by atomic disorders. The annealing procedure improves the AMR by reduction of the total resistivity. At the same time, we also find that the fourth-order term in the angular dependent resistivity becomes weaker when the disorder increases, resulting from increased scattering of the states around the band-crossing.
\end{abstract}

\section{Introduction}

The anisotropic magnetoresistance (AMR) effect, where the longitudinal resistivity in ferromagnetic alloy is dependent on the magnetization direction, was discovered by Thomson more than one hundred years ago \cite{1975-McGuire}. Its applications include advanced information storage and modern sensors \cite{2006-Popovic, 2018-Wang}. However, unrevealing its mechanism has been pondered \cite{1975-McGuire}. To explain the AMR, Fert and Campbell proposed the two-current model \cite{1968-Fert}, where spin-dependent scattering plays the central role. Residual resistivity and AMR of bulk NiFe alloy were calculated using the Kubo-Greenwood formula with the fully relativistic spin-polarized screened (KKR) method and the coherent potential approximation (CPA) \cite{2011-Ebert, 1997-Ebert}. Although the calculated residual resistivities are 30\textasciitilde40\% smaller than the measured values, the AMR ratios were in excellent agreement with the experiments in the whole composition percentage. Very recently, the topological variations of the bands with the magnetization directions were used to account for the AMR. The spin-orbital coupling (SOC) dependent gap between the anticrossing bands in the vicinity of the Fermi energy was the cause of the changing resistivity with respect to the magnetization direction \cite{2020-Zeng}. At the same time, the fourfold symmetric term of AMR was attributed to the relaxation time anisotropy due to the variation of the density of states near the Fermi energy under the rotation of the magnetization \cite{2022-Dai}. The fourfold AMR was also proposed to originate from the splitting of the 3\textit{d} states of the impurities owing to the SOI \cite{2020-Yahagi}. 

Nowadays, more details are included in the computation of AMR. The AMR at finite temperatures in hexagonal late transition metals and their alloys was calculated \cite{2020-Wagenknecht, 2020-Wagenknecht_2}, where phonon and spin fluctuation scattering were taken into account. The combination of both reproduced the experimental resistivity. Disorder can not only be introduced by phonon or magnon at finite temperatures but also by the atom occupations at the lattice. Atomic disorder is widely observed in alloys and compounds, especially materials with a B2 lattice or sublattice \cite{2006-Seeger}. In these materials, the measured AMR decreases with the annealing temperatures. Our previous work shows that the change in AMR is mainly due to the reduction of the total resistivity \cite{2017-Zhang}. It was usually attributed to the reduction of the grain boundary and the growth of the grain size \cite{2010-Toth, 2011-Wang}. The structure disorders can also be reflected in anisotropic magnetoresistance \cite{2012-Bruski}. There are always some complexities in explaining the AMR influenced by disorder. Disorder may increase the resistivity by enhancing the electron scattering and smearing the band structure. At the same time, the gaps formed in the anticrossing bands mentioned above will also change. In this work, we find that anti-site disorder induced mainly the broadening of the spectral function and the decrease in the conductivity, which is the main reason for the decrement of the AMR. In contrast, the difference in conductivity introduced by the rotation of the magnetization direction is not influenced by the disorder. Namely, the intrinsic origins of AMR are robust against disorder. We also find that the disorder increases the symmetry and leads to the decrement of the fourth-order term in AMR attached to the cubic symmetry.

The paper is organized as follows: In Sec. II, we present the details related to the calculation method and the parameters we used. In Sec. III, we analyze the factors affecting electronic transport based on the electronic structure and the Bloch spectral function. The paper is summarized in Sec. IV.
	
\section{Computational details}
	
In order to model the anti-site disorder in BCC FeCo alloy, the concentration of Fe at (0, 0, 0) was set to be \textit{x}, with the concentrations of Fe$_{(0.5, 0.5, 0.5)}$, Co$_{(0, 0, 0)}$, and Co$_{(0.5, 0.5, 0.5)}$ were set to be 1-\textit{x}, 1-\textit{x}, and \textit{x}, respectively. We defined the configuration entropy as $S=-k_B[x\ln{x}+(1-x)\ln{(1-x)}]$ to characterize the degree of anti-site disorder where $k_B$  is the Boltzmann constant.

The electronic structure calculation of Fe$_x$(Co)$_{1-x}$Co$_x$(Fe)$_{1-x}$ has been implemented within the SPR-KKR (spin-polarized relativistic Korringa-Kohn-Rostoker) program package \cite{2011-Ebert}. The anti-site disorder was treated within the coherent potential approximation (CPA) method \cite{1967-Soven}. We adopted the fully relativistic mode (REL) to include the relativistic effect, which introduces the spin-orbit coupling. Self-consistent potentials were obtained employing the Vosko-Wilk-Nussair (VWN) parametrization \cite{1980-Vosko} for the exchange-correlation functional in the local density approximation (LDA). The energy integration was performed on a semicircular path in the complex plane (GRID = 5) with 32 points of the \textit{E}-mesh. The atomic sphere approximation (ASA) was used for the potentials. $l_{max}$, the angular momentum expansion cutoff parameter, plays an essential role in the convergence of KKR calculations \cite{2014-Alam}. According to the previous works \cite{2013-Mankovsky, 2016-Ayaz_Khan} and our validation, $l_{max}=3$ is sufficient for $d$-systems. To obtain the converged calculation, we used 7.2 $\times$ $10^{7}$ \textit{k} points in the full Brillouin zone. The Bloch spectral function (BSF) is applied to visualize the electronic structure of alloys, which represents the spectral weight distribution of electrons in energy-momentum space \cite{2016-Chadova}.

The Kubo linear response formalism was adopted for dealing with electronic transport in metallic systems. Based on the formalism, using the independent electron approximation and the thermal limit (T = 0 K), one arrives at the Kubo-Středa equation for the conductivity tensor \cite{1982-Streda}

\begin{equation}
	\begin{split}
		\sigma_{\mu v} & =\frac{\hbar}{4 \pi V} \operatorname{Tr}\langle\hat{J}_\mu(G^{+}-G^{-}) \hat{J}_v G^{-} -\hat{J}_\mu G^{+} \hat{J}_v(G^{+}-G^{-})\rangle_c \\ 
		& +\frac{e}{4 \pi i V} \operatorname{Tr}\langle(G^{+}-G^{-})(\hat{r}_\mu \hat{J}_v-\hat{r}_v \hat{J}_\mu)\rangle_c.
		\label{eq1}
	\end{split}
\end{equation}

Here $\hat{r}$ and $\hat{J}$ are the position and current density operators, respectively. $\mu$ and $v$ denote the Cartesian coordinates. $V$ is the unit cell volume. The retarded $\left(G^{+}\right)$ and advanced $\left(G^{-}\right)$ Green functions represent the electronic structure of the system. $\langle\cdots\rangle_c$ indicates a configurational average performed within the CPA method.
	
\section{Results and discussions}
\subsection{The lattice constant and the density of states}

\begin{figure}[htb]
	\centering
	
	\includegraphics[width=0.32\linewidth]{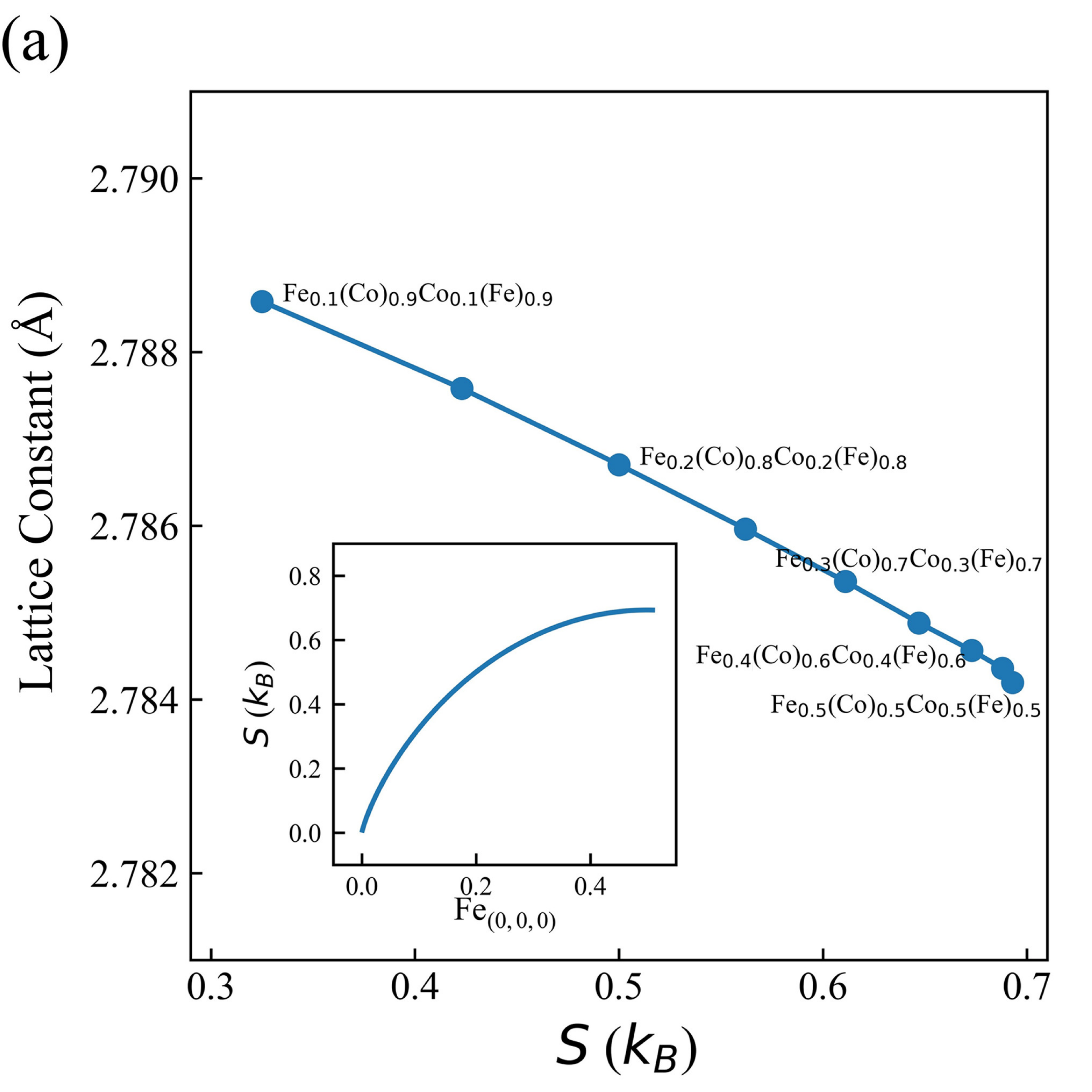}
	\includegraphics[width=0.32\linewidth]{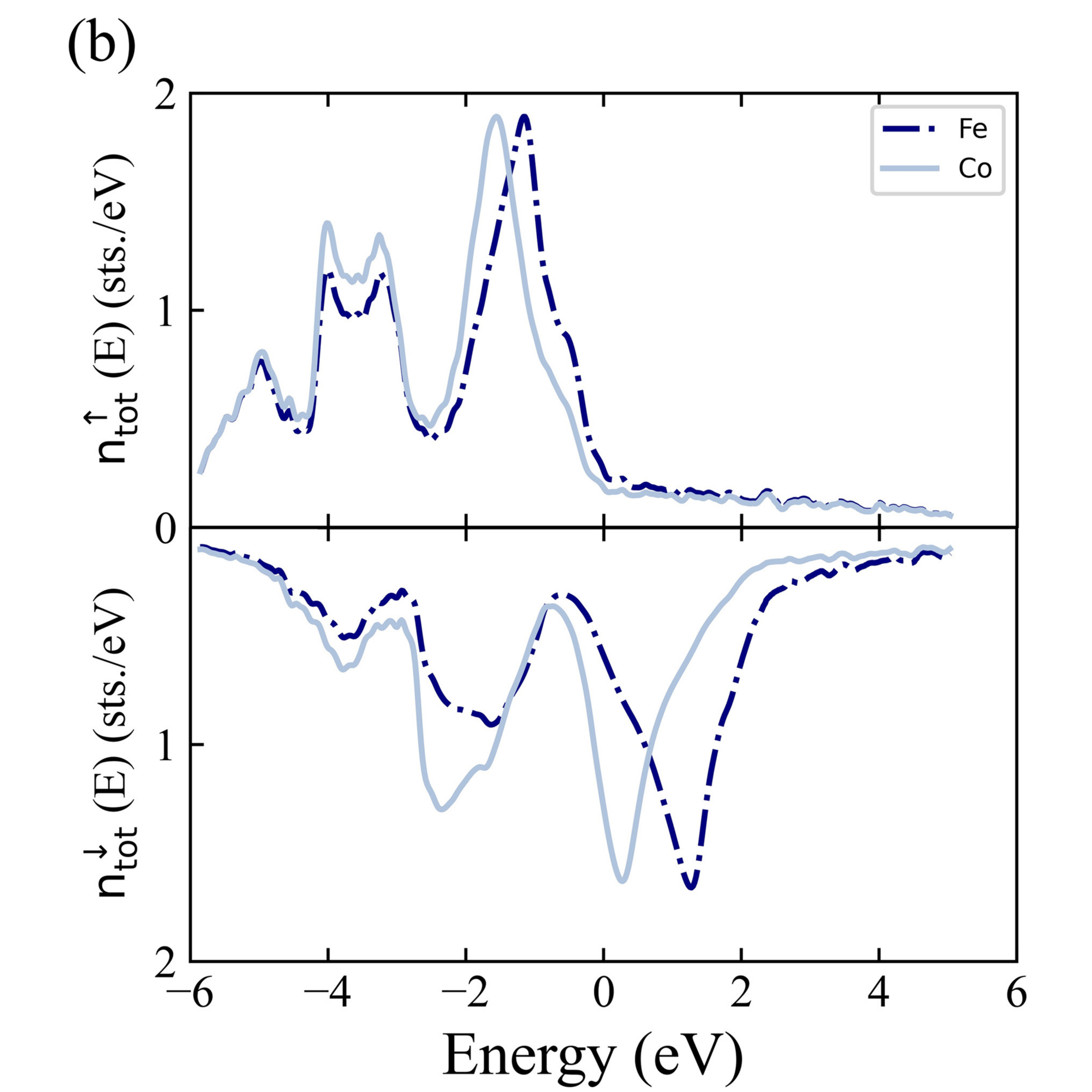}
	\includegraphics[width=0.32\linewidth]{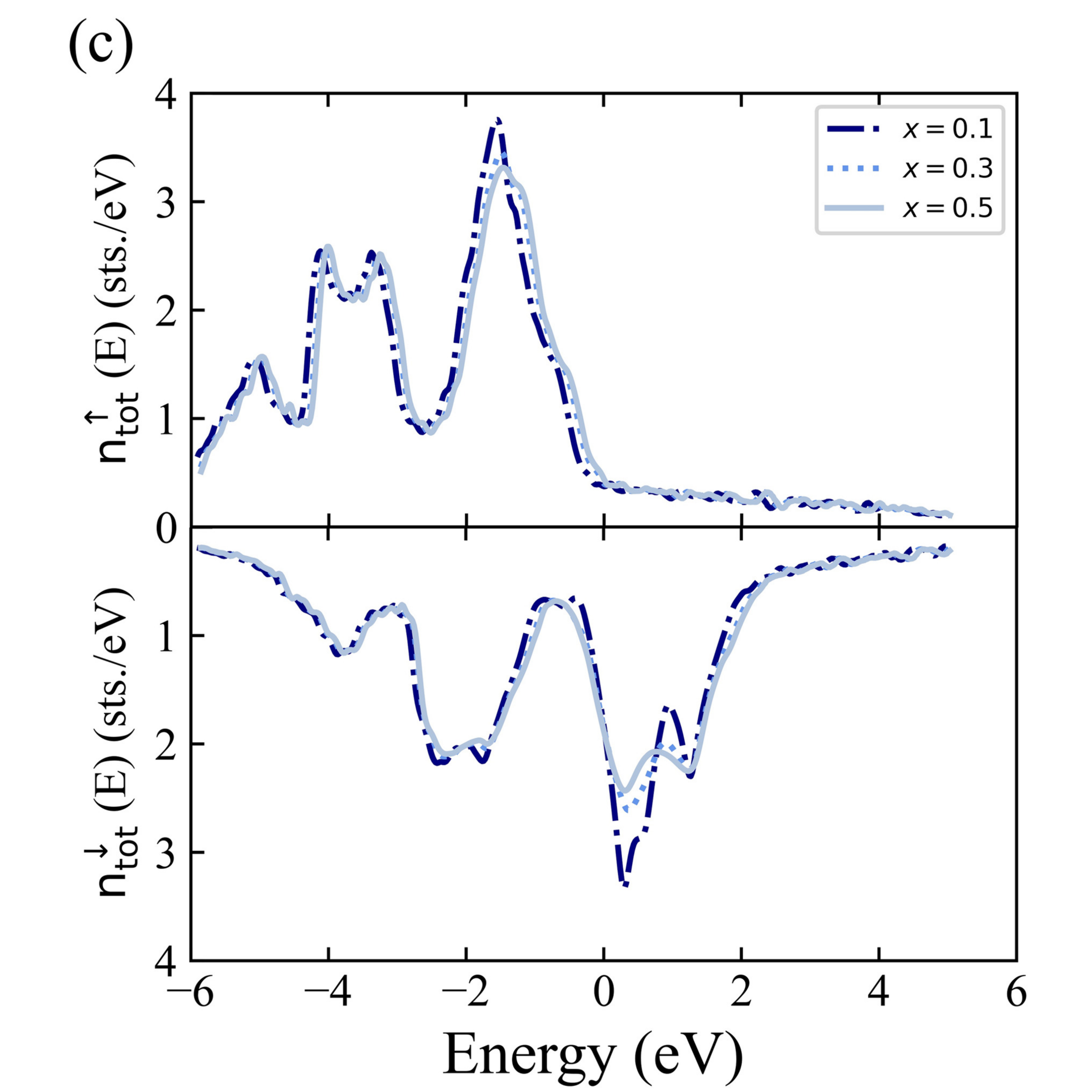}
	\caption{(a) The lattice constant decreases with the increase of $S$. The inset is the relationship between $S$ and the concentration of Fe$_{(0, 0, 0)}$. (b) The DOS projected on $\mathrm{Fe}$ and $\mathrm{Co}$ in Fe$_{0.5}$(Co)$_{0.5}$Co$_{0.5}$(Fe)$_{0.5}$. (c) The densities of states of Fe$_x$(Co)$_{1-x}$Co$_x$(Fe)$_{1-x}$ alloys with different degrees of anti-site disorder.}
	\label{fig1}
\end{figure}

The entropy (\textit{S}) grows monotonically from 0 $k_B$ to 0.693 $k_B$ with the increasing concentration of  Fe$_{(0, 0, 0)}$. It reaches the maximum when the concentrations of Fe and Co at the same site become identical. For this reason, only the name of Fe$_x$(Co)$_{1-x}$Co$_x$(Fe)$_{1-x}$ alloy is used to indicate the degree of anti-site disorder in the subsequent part.

The lattice constant decreases with the increasing degree of disorder, the same as the results of Kašpar et al. \cite{1983-Kaspar} and Jacob et al. \cite{2020-Jacob}. Moreover, Kašpar et al. suggested that the change in the lattice constant is due to some occupied anti-bonding majority-spin molecular orbitals becoming partially unoccupied in the transition from order to disorder \cite{1983-Kaspar}. Such a change in the lattice constant will induce variations in the DOS as well as the magnetic moment.

The average moment slightly depends on the anti-site disorder, rising from 2.11 $\mu_{B}$ to 2.13 $\mu_{B}$, in agreement with the conclusion of Spooner et al. \cite{1972-Spooner}. Because the Co atom is more electronegative and has a smaller exchange splitting than Fe, the minority-spin states of Co lie at lower energy than the Fe states \cite{2002-Kulkova}. Consequently, the DOS at the Fermi energy comes mainly from the spin-down electrons of Co, as shown in FIG. \ref{fig1} (b), so the magnetic moment should primarily originate from cobalt. 

Regarding the DOSs with different degrees of anti-site disorder, which are exhibited in FIG. \ref{fig1} (c), the ferromagnetically ordered alloys have only slight changes in the DOS of the spin-up electrons. While for the spin-down parts, substantial decreases in the peaks above the $E_{\rm{F}}$ smoothen the shape of DOS, meaning a weaker localization. The sharp contrast between the DOS of majority and minority electrons is consistent with the conclusions of Kulkova et al. \cite{2002-Kulkova}. The flip of the majority-occupied states to the minority-unoccupied ones will decrease the spin moment of the alloy, compensating for the effect of localization.

For ferromagnet, the essential details lie in the minority spin system \cite{1984-Schwarz}, so it can be expected that the reduction of the DOS at the $E_{\rm{F}}$ due to anti-site disorder will increase the resistivity of the alloys.

\subsection{Resistivity and its angular dependence}

\begin{figure}[htb]
	\centering
	\includegraphics[width=0.4\linewidth]{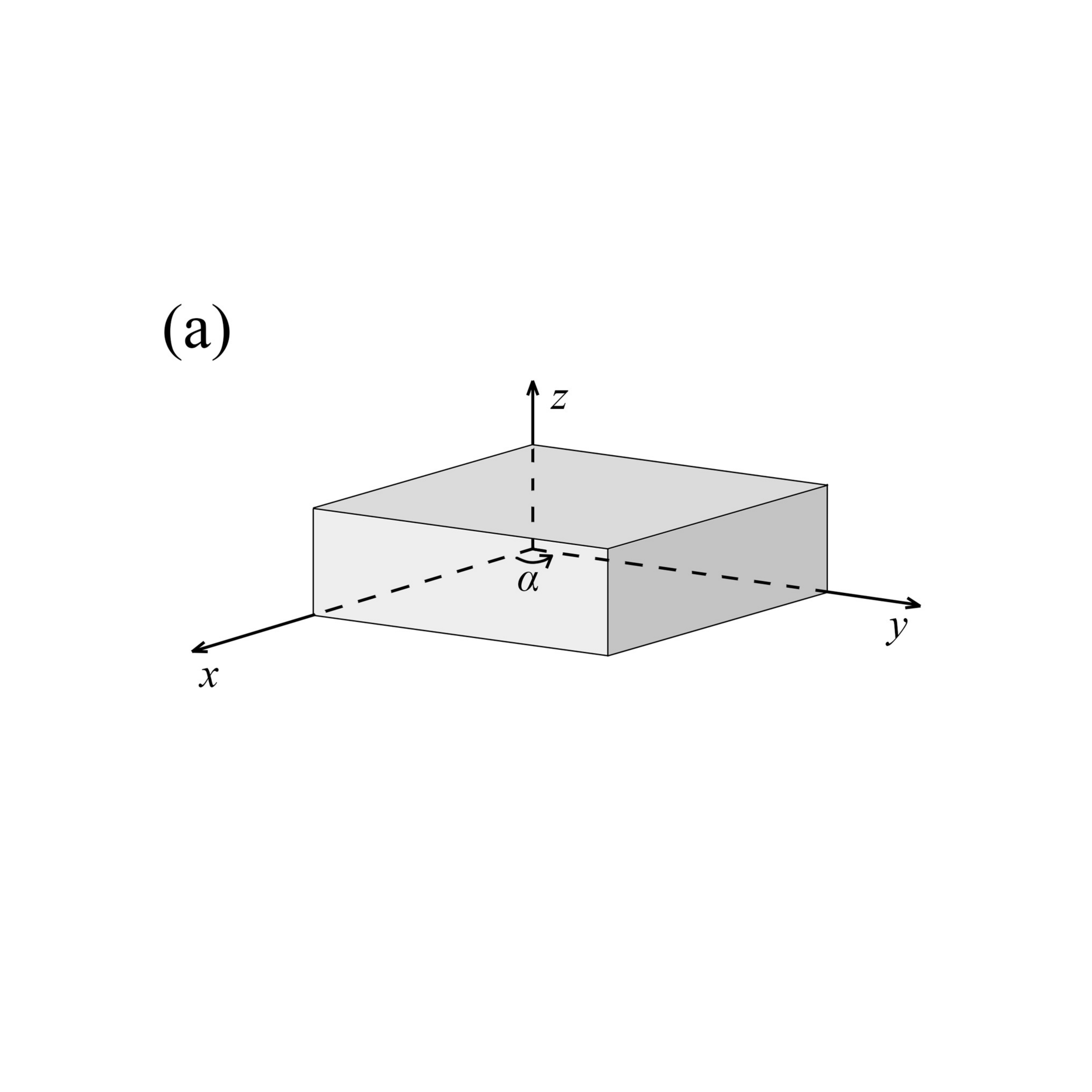}
	\includegraphics[width=0.4\linewidth]{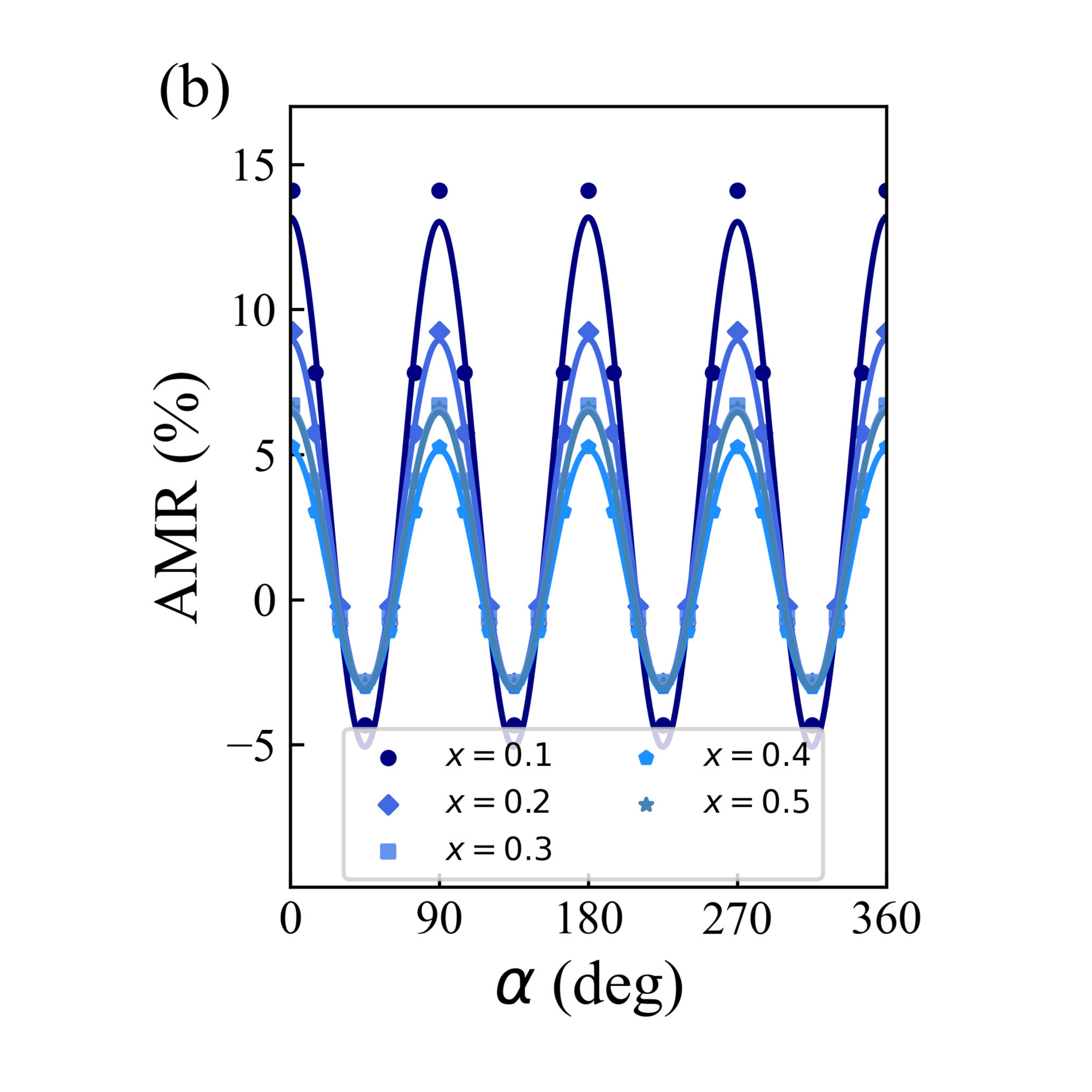}
	
	\includegraphics[width=0.4\linewidth]{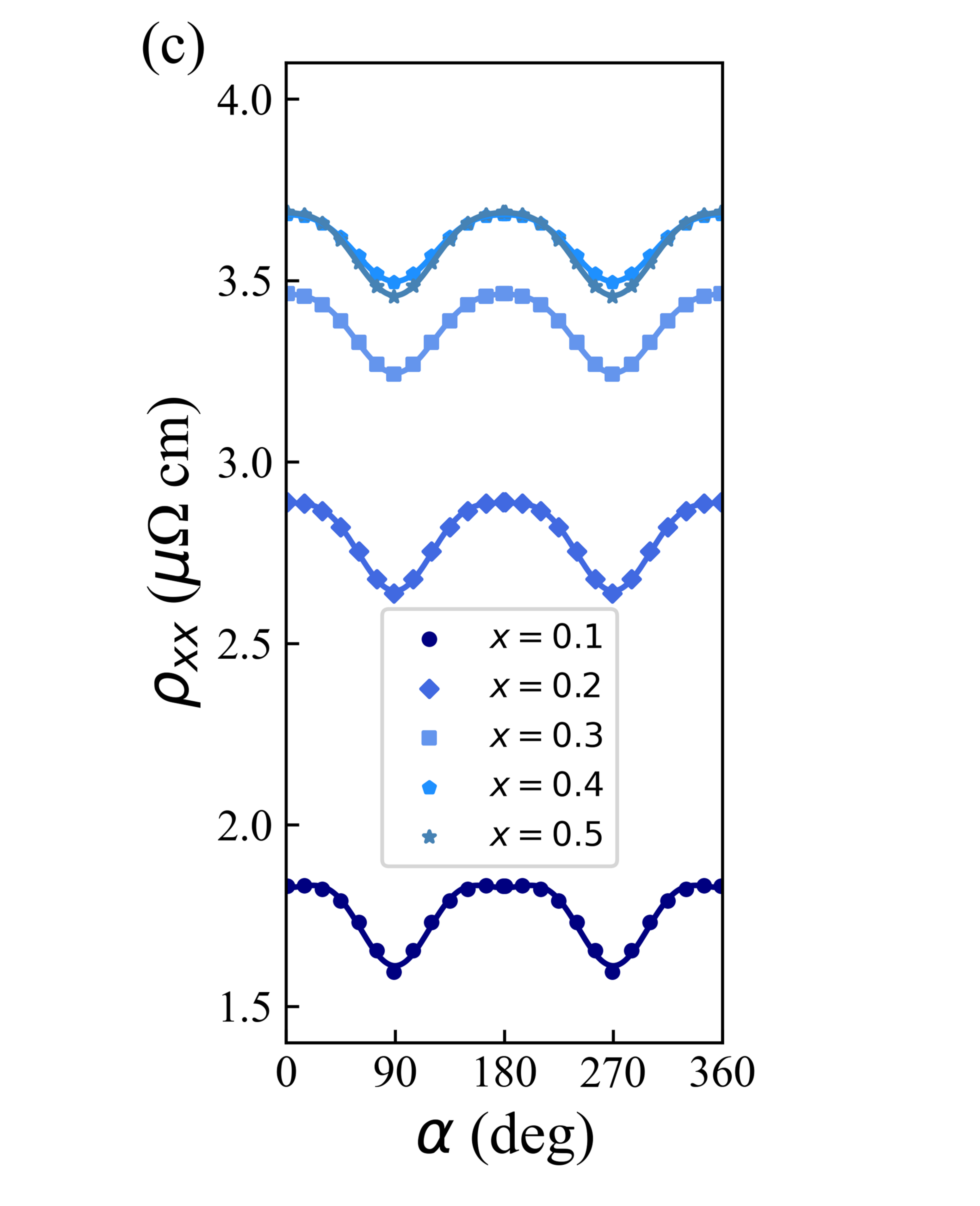}
	\includegraphics[width=0.4\linewidth]{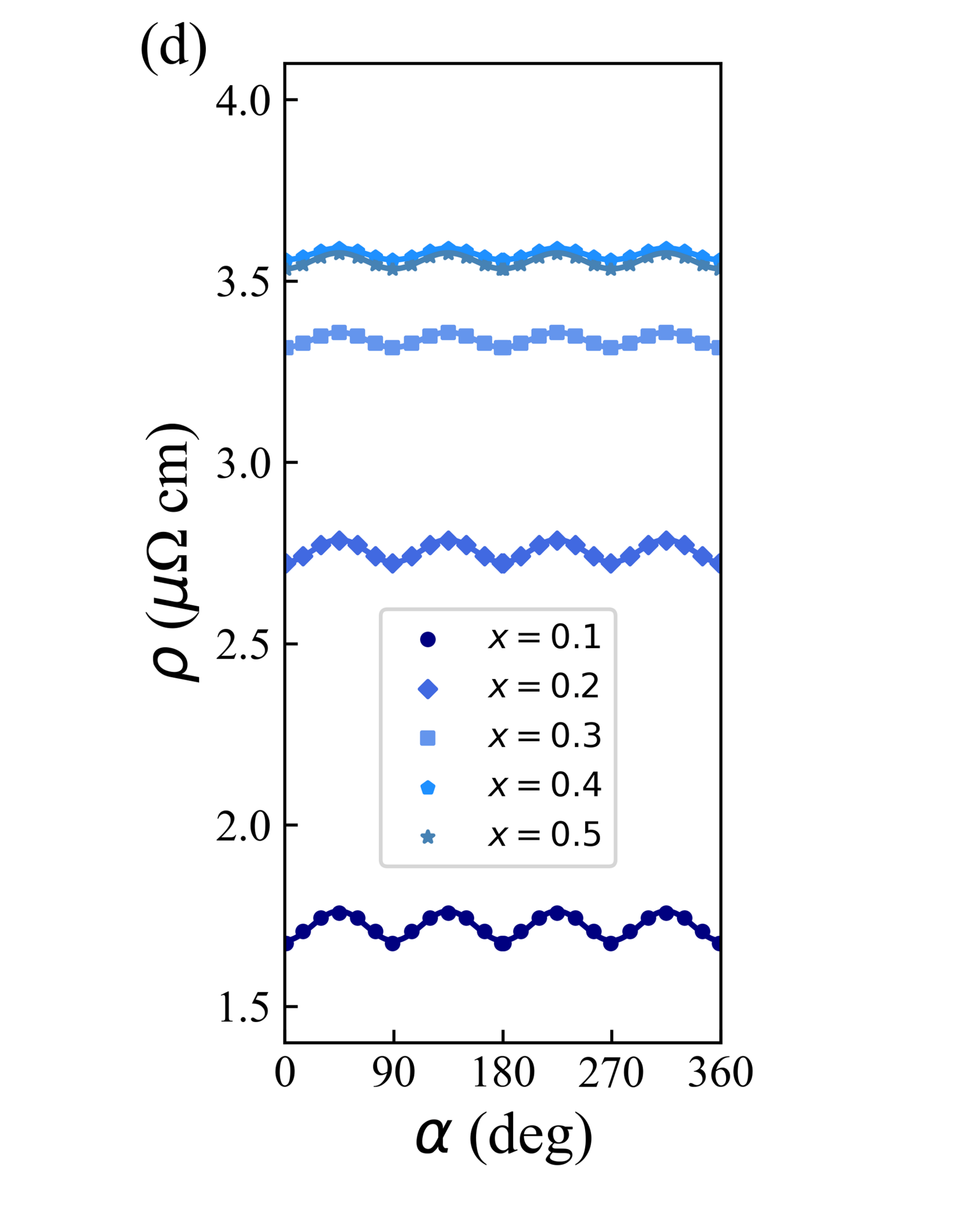}
	\caption{(a) The definition of angle $\alpha$ in Cartesian coordinates. The calculated AMR (b), diagonal resistivity $\rho_{x x}$ (c), and average resistivity $\rho$ (d) of Fe$_x$(Co)$_{1-x}$Co$_x$(Fe)$_{1-x}$ alloys. The curves in (c) and (d) are fitted to Eq. (2) and (3), respectively.}
	\label{fig2}
\end{figure}

Since the resistivity varies with the magnetization direction, $\mathrm{AMR}=\frac{\left(\rho_{\|}-\rho_{\perp}\right)}{\left(\frac{1}{3} \rho_{\|}+\frac{2}{3} \rho_{\perp}\right)} \times 100 \%$ is defined to characterize the anisotropic magnetoresistance, where $\rho_{\|}\left(\rho_{\perp}\right)$ is the resistivity when the magnetization is parallel (perpendicular) to the current direction. According to the transformation law for the second rank tensors, we obtained $\rho_{\|}=\cos ^2 \alpha \rho_{x x}+\cos \alpha \sin \alpha \rho_{x y}+\cos \alpha \sin \alpha \rho_{y x}+\sin ^2 \alpha \rho_{y y}$ and $\rho_{\perp}=$ $\sin ^2 \alpha \rho_{x x}-\cos \alpha \sin \alpha \rho_{x y}-\cos \alpha \sin \alpha \rho_{y x}+\cos ^2 \alpha \rho_{y y}$ from the resistivity tensors, where $\rho_{ij}$ is the component of the resistivity tensor and $\alpha$ is the angle between the magnetization direction and the $x$-axis. The AMR values for different anti-site disordered Fe$_x$(Co)$_{1-x}$Co$_x$(Fe)$_{1-x}$ alloys are shown in FIG. \ref{fig2} (b), and the solid curves represent the fitting results. As anti-site disorder increases, the AMR decreases from 14.1\% to 5.3\% dramatically, meaning the alloy becomes more isotropic than before.

The longitudinal and average resistivity $\rho_{x x}$ and $\rho$ can be written as series expansions with respect to $\alpha$ based on the phenomenological model \cite{1975-McGuire}. Since we consider the situation when the magnetic moment rotates in the $x y$-plane, the expansions can be simplified as

\begin{equation}
	\rho_{x x}=A+B \cos 2 \alpha-D \cos 4 \alpha,
	\label{eq2}
\end{equation}

\begin{equation}
	\rho =\frac{(\rho_{x x}+\rho_{y y}+\rho_{z z})}{3} =(A-\frac{B}{3}-\frac{D}{3}+\frac{\mathrm{E}}{12})-(\frac{2 D}{3}+\frac{\mathrm{E}}{12}) \cos 4 \alpha.
	\label{eq3}
\end{equation}

$\rho_{x x}$ and $\rho$ are fitted to the magnetic moment angle $\alpha$ and shown in FIG. \ref{fig2} (c) and (d). The angular dependence of $\rho_{x x}$ is in good agreement with the results of Zeng et al. \cite{2020-Zeng}. In addition to the change in magnetization direction, which affects the resistivity, the anti-site disorder also significantly affects $\rho_{x x}$ and $\rho$. The comparison of the five Fe$_x$(Co)$_{1-x}$Co$_x$(Fe)$_{1-x}$ alloys in FIG. \ref{fig2} (c) and (d) can lead to the conclusion that anti-site disorder causes enhanced scattering and a substantial increase in resistivity. According to the experiments of Freitas et al., the FeCo resistivity increases from 2.09 $\mu\Omega\rm{cm}$ to 3.24 $\mu\Omega\rm{cm}$ with the increasing disorder by varying the quenching temperatures \cite{1988-Freitas}, which clearly connects the atomic disorder effect on the electronic transport.

\begin{figure}[htb]
	\centering
	\includegraphics[width=0.4\linewidth]{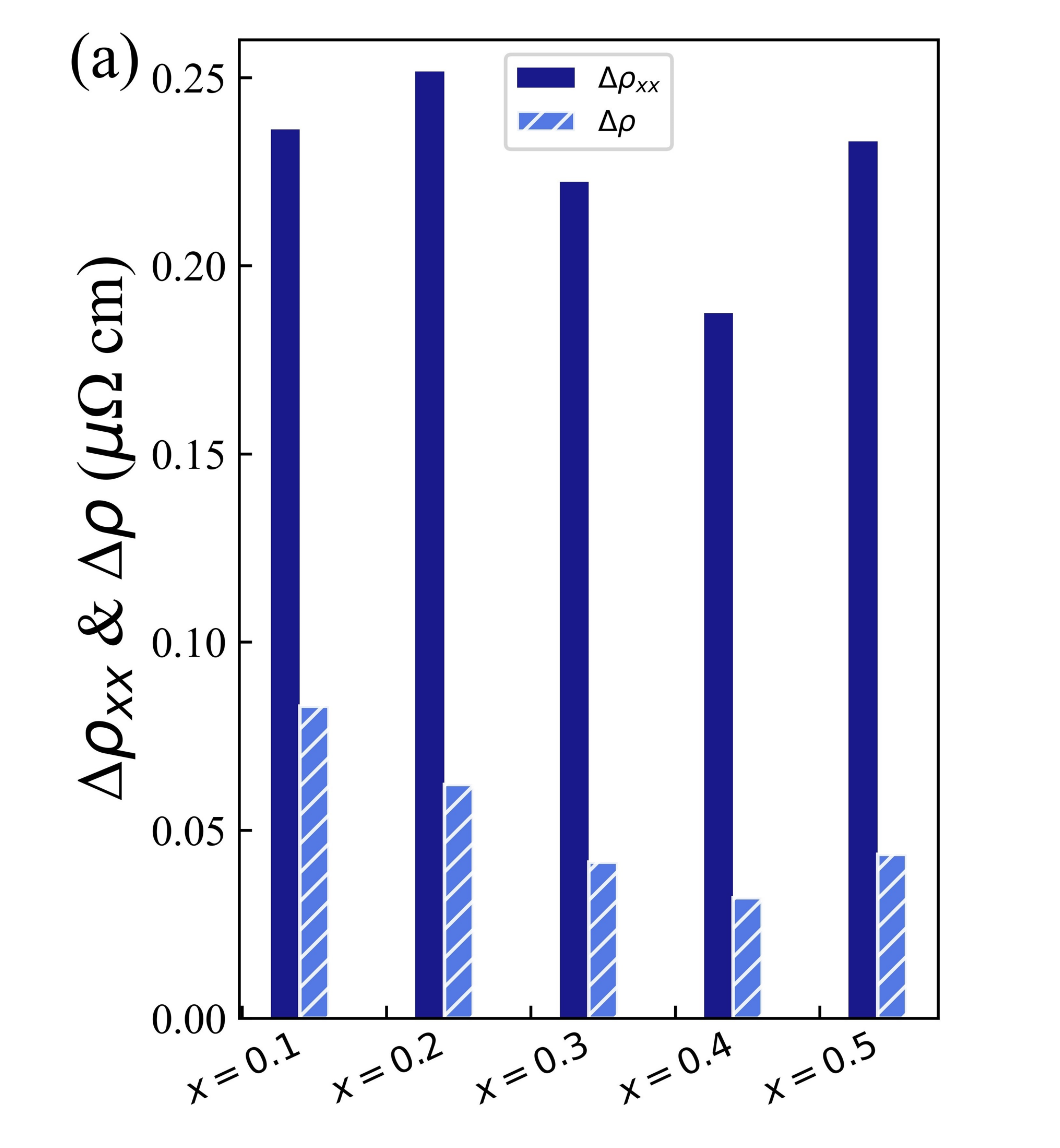}
	\includegraphics[width=0.4\linewidth]{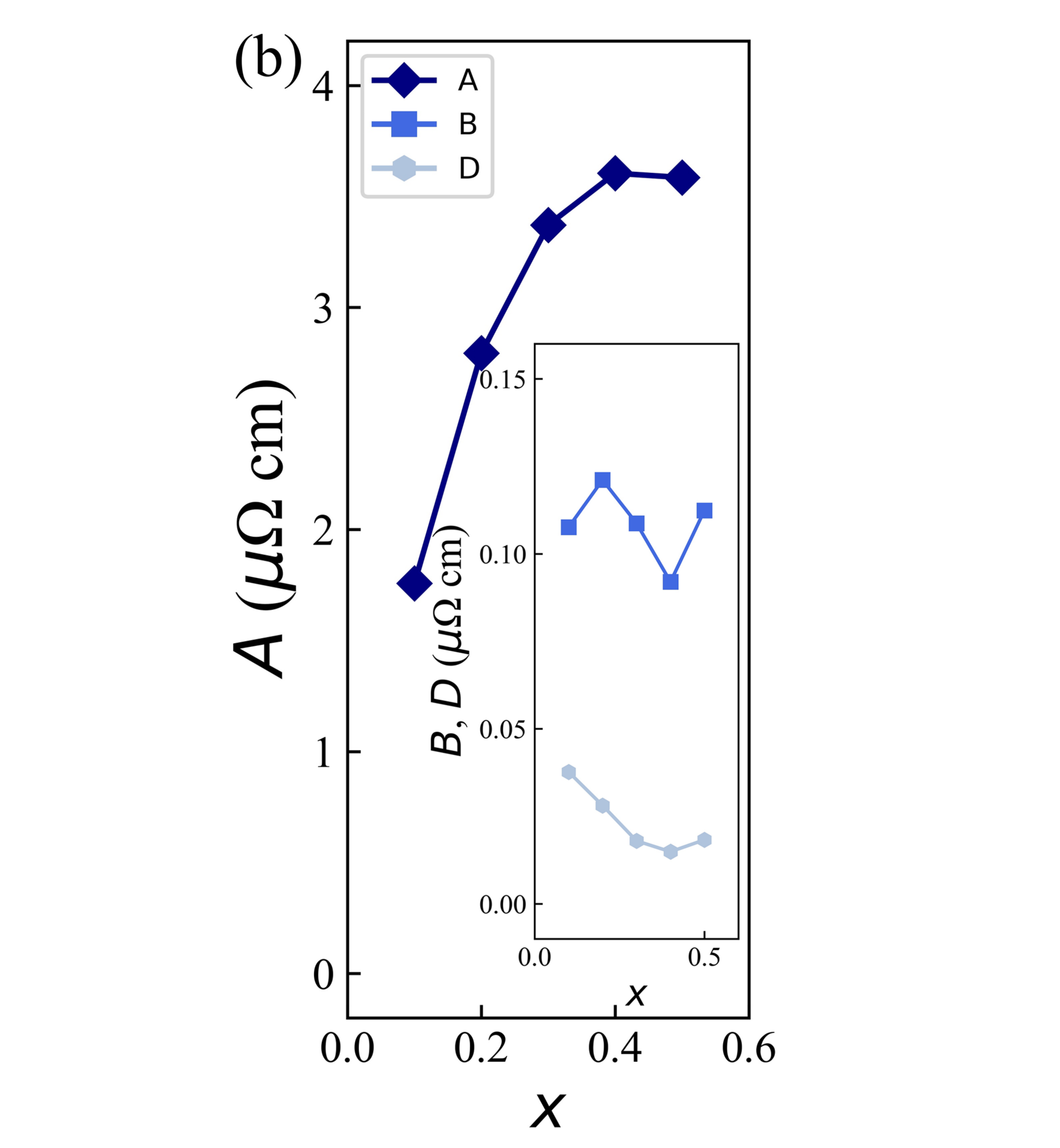}
	\caption{(a) The $\Delta \rho$ and $\Delta \rho_{x x}$ of different disordered Fe$_x$(Co)$_{1-x}$Co$_x$(Fe)$_{1-x}$ alloys. (b) The coefficients of the phenomenological expression of $\rho_{x x}$ as a function of the concentration of Fe$_{(0, 0, 0)}$.}
	\label{fig3}
\end{figure}

To find out the reason for the decrease in AMR with the anti-site disorder, we define $\Delta \rho=\rho_{\max }-\rho_{\min }$ and $\Delta \rho_{x x}=\rho_{x x, \max }-\rho_{x x, \min }$. It can be observed in FIG. \ref{fig3} (a) that $\Delta \rho_{x x}$ fluctuates around 0.22 $\mu\Omega\rm{cm}$, in agreement with the data of Zeng et al. \cite{2020-Zeng} and Wimmer et al. \cite{2018-Wimmer}. The values of $\Delta \rho_{x x}$ and $\Delta \rho$ are much smaller than $\rho_{x x}$ and $\rho$, however, indicating that the difference in AMR between different disordered systems is mainly a result of the resistivity increase caused by disorder rather than the change in $\rho_{x x}$ and $\rho$ due to the rotation of the magnetization direction. The enhancement of anti-site disorder also dramatically raises the value of the constant term $A$. The $\cos 4 \alpha$ term's coefficient $D$, which characterizes the single crystal anisotropic magnetoresistance, reduces remarkably with the degree of disorder, as shown in FIG. \ref{fig3} (b).

The vertex correction is crucial for discussing the scattering of impurities in electronic transport \cite{2008-Tulip, 2020-Sipr}, which corresponds to the scattering-in term in the Boltzmann equation \cite{1985-Butler, 2019-Turek}. However, the effect of the vertex correction is much weaker if the states at the Fermi energy have $d$ properties primarily \cite{1994-Banhart, 2004-Turek, 2015-Ebert}. In KKR-CPA calculation, the vertex correction of conductivity obtained based on the Kubo linear response theory is the difference between correlated and uncorrelated configurational averages

\begin{equation}
	\sigma_{\mu \nu}^{\rm{V C}}=\left\langle\hat{O}_\mu \hat{G}^{+} \hat{J}_\nu \hat{G}^{-}\right\rangle_c-\left\langle\hat{O}_\mu \hat{G}^{+}\right\rangle_c\left\langle\hat{J}_\nu \hat{G}^{-}\right\rangle_c.
	\label{eq3}
\end{equation}

\begin{figure}[htb]
	\centering
	\includegraphics[width=0.4\linewidth]{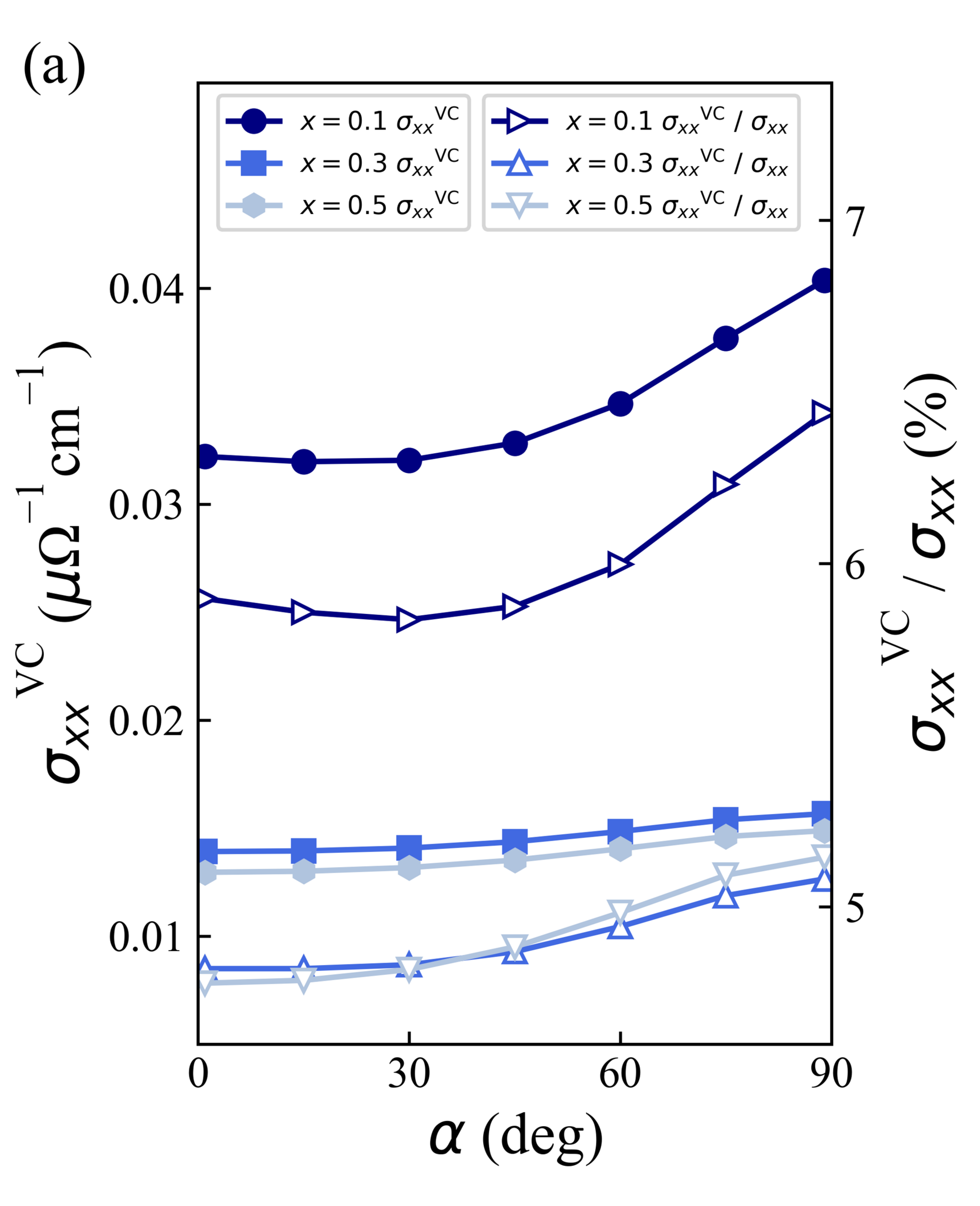}
	\includegraphics[width=0.4\linewidth]{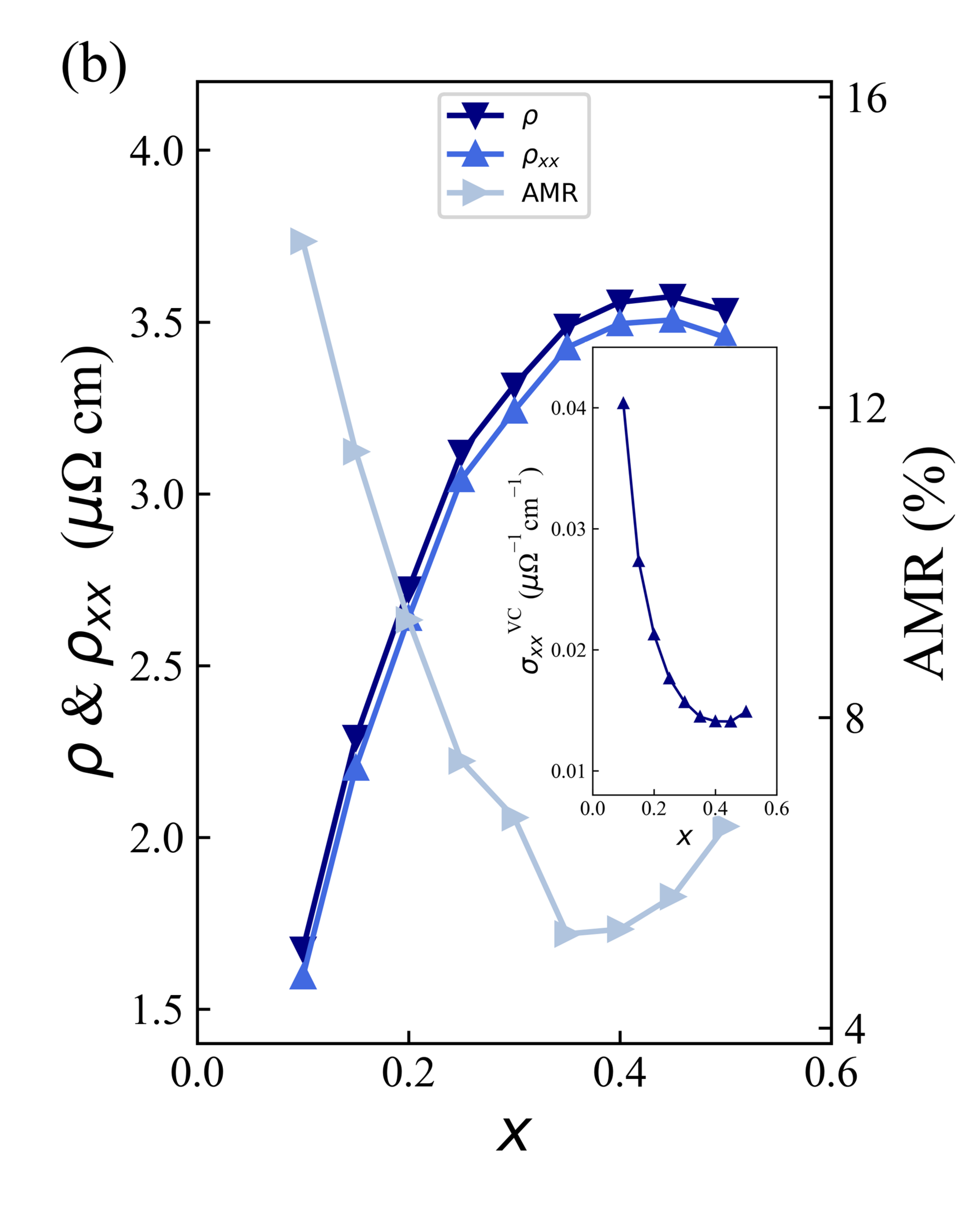}
	\caption{(a) The vertex correction of the longitudinal conductivity $\sigma_{x x}^{\mathrm{VC}}$ and $\sigma_{x x}^{\mathrm{VC}}$ / $\sigma_{x x}$ as a function of $\mathbf{M}$. (b) Dependencies of $\rho$, $\rho_{x x}$, $\rm{AMR}$, and $\sigma_{x x}^{\mathrm{VC}}$ on the anti-site disorder.}
	\label{fig4}
\end{figure}

FIG. \ref{fig4} (a) shows the vertex correction of the longitudinal conductivity, $\sigma_{x x}^{\mathrm{VC}}$, which strongly depends on the direction of magnetization. From the numerical point of view, however, $\sigma_{x x}^{\mathrm{VC}}$ accounts for only 5\% to 7\% of $\sigma_{x x}$ in the Fe$_x$(Co)$_{1-x}$Co$_x$(Fe)$_{1-x}$ alloy since its Fermi level is mainly occupied by the 3$d$ electrons. In a semiclassical picture, when $\mathbf{M}$ gradually approaches the $x$-axis, the scattering enhancement causes the scattering events experienced by different electrons to tend to be the same. Therefore, the difference between correlated and uncorrelated configurational averages, i.e., the vertex correction $\sigma_{x x}^{\mathrm{VC}}$, decreases. The higher the degree of anti-site disorder, the stronger the scattering, which leads to less sensitivity to the rotation of $\mathbf{M}$ and a smaller value of $\sigma_{x x}^{\mathrm{VC}}$, as shown in FIG. \ref{fig4} (a) and (b). Moreover, the variation of $\sigma_{x x}^{\mathrm{VC}}$ due to the rotation of $\mathbf{M}$ is much smaller than that introduced by the anti-site disorder, confirming that the disorder-induced scattering is more substantial in the Fe$_x$(Co)$_{1-x}$Co$_x$(Fe)$_{1-x}$ alloy.

Interestingly, the maximal values of $\rho$ and $\rho_{x x}$ do not occur at $x=0.5$, but where the concentration of $\mathrm{Fe}_{(0,0,0)}$ is around 45\%, as shown in FIG. \ref{fig4} (b). Therefore, the differences of $\rho$ and $\rho_{x x}$ between $x=0.5$ and $x=0.4$ in FIG. \ref{fig2}  are negligible. The non-monotonic increment of the conductivity with the disorder depends on the detailed electronic structure, which will be seen in the following discussions.

\subsection{The Bloch spectral function}

\begin{figure}[htb]
	\centering
	\includegraphics[width=0.8\linewidth]{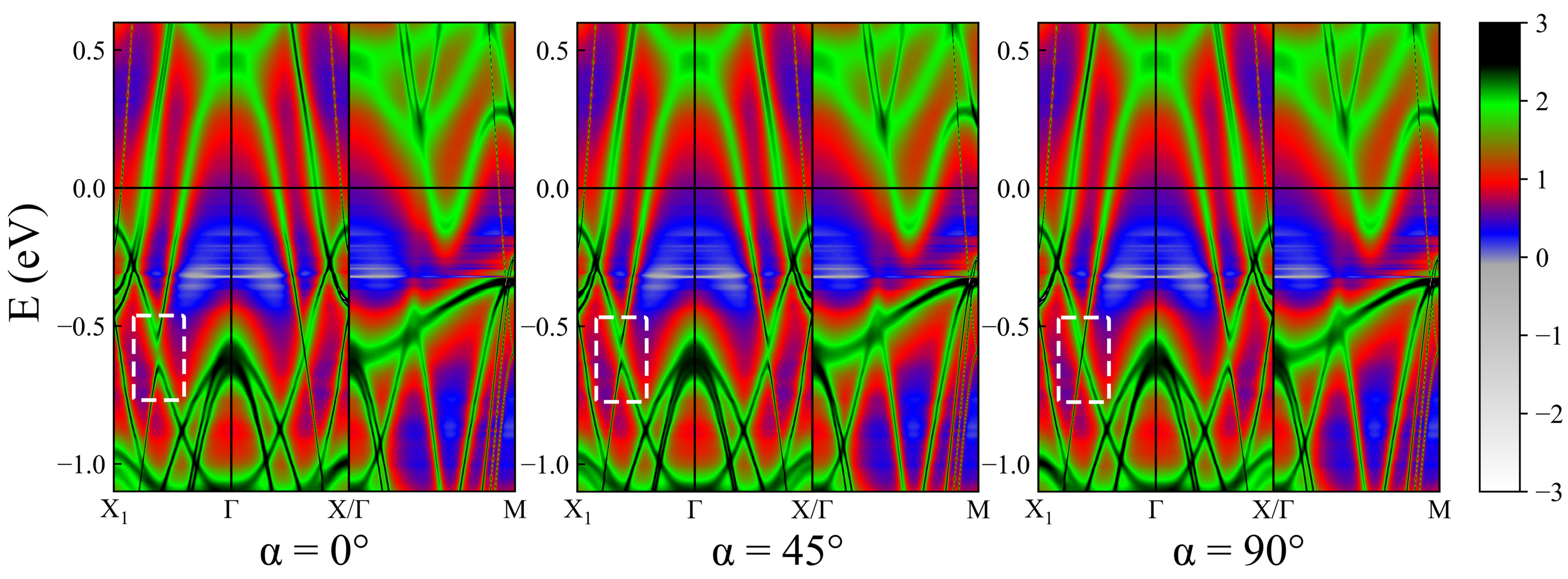}
	\caption{The total BSF of Fe$_{0.1}$(Co)$_{0.9}$Co$_{0.1}$(Fe)$_{0.9}$ when $\alpha=0^{\circ}$ (left), $\alpha=45^{\circ}$ (middle), and $90^{\circ}$ (right). The Fermi level is set at 0 eV.}
	\label{fig5}
\end{figure}

The AMR comes from the relative variation of longitudinal resistivity $\rho_{x x}$ and total resistivity $\rho$ when the magnetic moment rotates. The two values may have various dependencies on the magnetic moment for different alloys. The electronic reason for the difference can be observed from the Bloch spectral function (BSF), with the color bar indicating the magnitude of $\log _{10} A(\vec{k}, E)$. The absence of translational symmetry due to random occupation of the lattice sites makes the Bloch state no longer an eigenstate of the system. As a result, the system does not have the well-defined dispersion found in ordered systems. The smeared bands in the BSF caused by scattering imply a reduction of the electron mean free path \cite{2015-Zhang}. At the same time, when the relativistic effect is taken into account, the spin is no longer a good quantum number. The spin-up and down states are mixed when the magnetization directions vary in space. \cite{1997-Ebert}. We show the total Bloch spectral function in FIG. \ref{fig5}, where $\mathrm{X}_1$ is the high symmetry point X in the [010] direction.

The disorder-induced smearing of the state at the Fermi level remains under different magnetization directions. In addition, we find no difference in the DOS at the $E_{\mathrm{F}}$ for the different magnetization directions. However, the band-crossing in the $\mathrm{X}-\Gamma$ path about 0.5 eV below the Fermi level gradually opens. The band-crossing responds to the AMR in the ordered compounds, as recently proposed by Zeng et al. \cite{2020-Zeng}. Simultaneously, a change of degeneracy occurs near the high symmetry point $\mathrm{X}_1$ 0.4 eV below the $E_{\mathrm{F}}$ in the $\mathrm{X}_1-\Gamma$ path, as shown in FIG. \ref{fig5}. Although they are slightly away from the $E_{\mathrm{F}}$, because of the smeared band, it can be qualitatively assumed that the alteration of the magnetization direction affects the electronic transport mainly in the form of intrinsic mechanisms \cite{2020-Zeng}. There is no difference between the BSFs when $\alpha=0^{\circ}$ and $90^{\circ}$, except that the band crossing/anticrossing positions are interchanged due to the symmetry of the crystal. For $\alpha=45^{\circ}$, however, the crossing opens on both $\mathrm{X}_1-\Gamma$ and $\mathrm{X}-\Gamma$ paths. In other words, the band-crossing caused by the change of $\mathbf{M}$ has the same fourfold symmetry as what contributes to the $\cos 4 \alpha$ term of AMR. 

\begin{figure}[htb]
	\centering
	\includegraphics[width=0.8\linewidth]{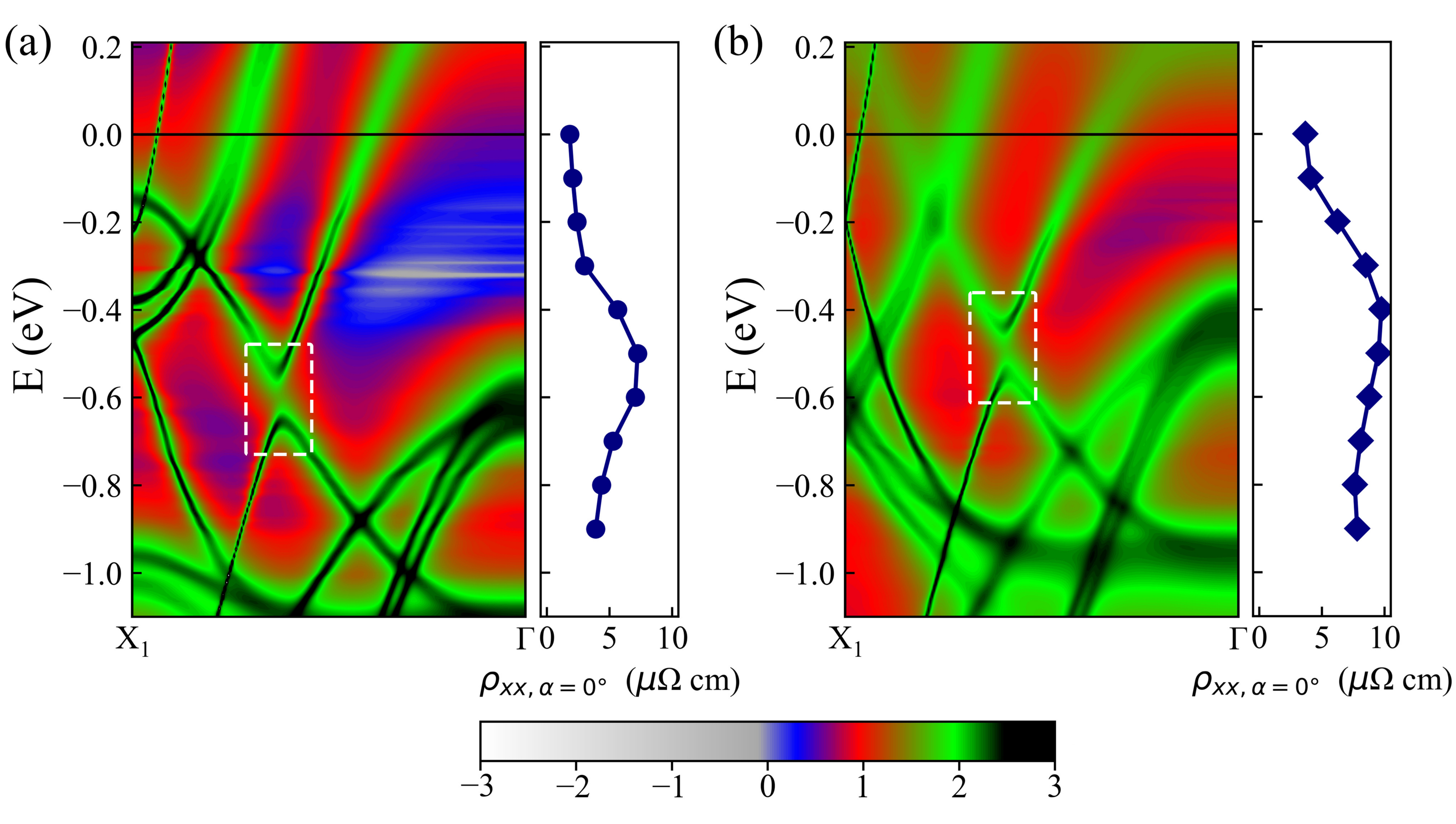}
	\caption{The total BSF of Fe$_{0.1}$(Co)$_{0.9}$Co$_{0.1}$(Fe)$_{0.9}$ (a) and Fe$_{0.5}$(Co)$_{0.5}$Co$_{0.5}$(Fe)$_{0.5}$ (b) at $\alpha=0^{\circ}$. The Fermi level is set at 0 eV in the BSF. $\rho_{x x, \alpha=0^{\circ}}$ is plotted as a function of energy.}
	\label{fig6}
\end{figure}

To examine the specific effect of changes in the band structure on the electronic transport, we manually alter the positions of $E_{\mathrm{F}}$ of Fe$_x$(Co)$_{1-x}$Co$_x$(Fe)$_{1-x}$ with $x=0.1$ and $x=0.5$, from which the resistivity contributed from the corresponding positions can be obtained. Thereby, one can further probe the relationship between the band crossing/anticrossing caused by the variation of $\mathbf{M}$ and the angular dependence of resistivity. In FIG. \ref{fig6}, the band-anticrossing of Fe$_{0.1}$(Co)$_{0.9}$Co$_{0.1}$(Fe)$_{0.9}$ appears around -0.6 eV, and $\rho_{x x}$ is notably enhanced around there. While for Fe$_{0.5}$(Co)$_{0.5}$Co$_{0.5}$(Fe)$_{0.5}$, the band-anticrossing occurs around -0.5 eV due to the lowered $E_{\mathrm{F}}$ caused by the higher anti-site disorder. The energy corresponding to the peak of $\rho_{x x}$ is accordingly lifted to -0.5 eV. On the other hand, since Fe$_{0.5}$(Co)$_{0.5}$Co$_{0.5}$(Fe)$_{0.5}$ is much more disordered than Fe$_{0.1}$(Co)$_{0.9}$Co$_{0.1}$(Fe)$_{0.9}$, its $\rho_{x x}$ curve with energy is much smoother and numerically higher than that of Fe$_{0.1}$(Co)$_{0.9}$Co$_{0.1}$(Fe)$_{0.9}$.

From the comparison of Fe$_x$(Co)$_{1-x}$Co$_x$(Fe)$_{1-x}$ with $x=0.1$ and $x=0.5$ in FIG. \ref{fig6}, in addition to the change in the band structure, the smearing is significantly strengthened in the BSF of Fe$_{0.5}$(Co)$_{0.5}$Co$_{0.5}$(Fe)$_{0.5}$, indicating the pronounced scattering enhancement. With the results of vertex correction, $\sigma_{x x}^{\mathrm{VC}}$ decreases with the augmentation of $S$, confirming that anti-site disorder drastically enhances the scattering, leading to a larger resistivity.

\begin{figure}[htb]
	\centering
	\includegraphics[width=0.4\linewidth]{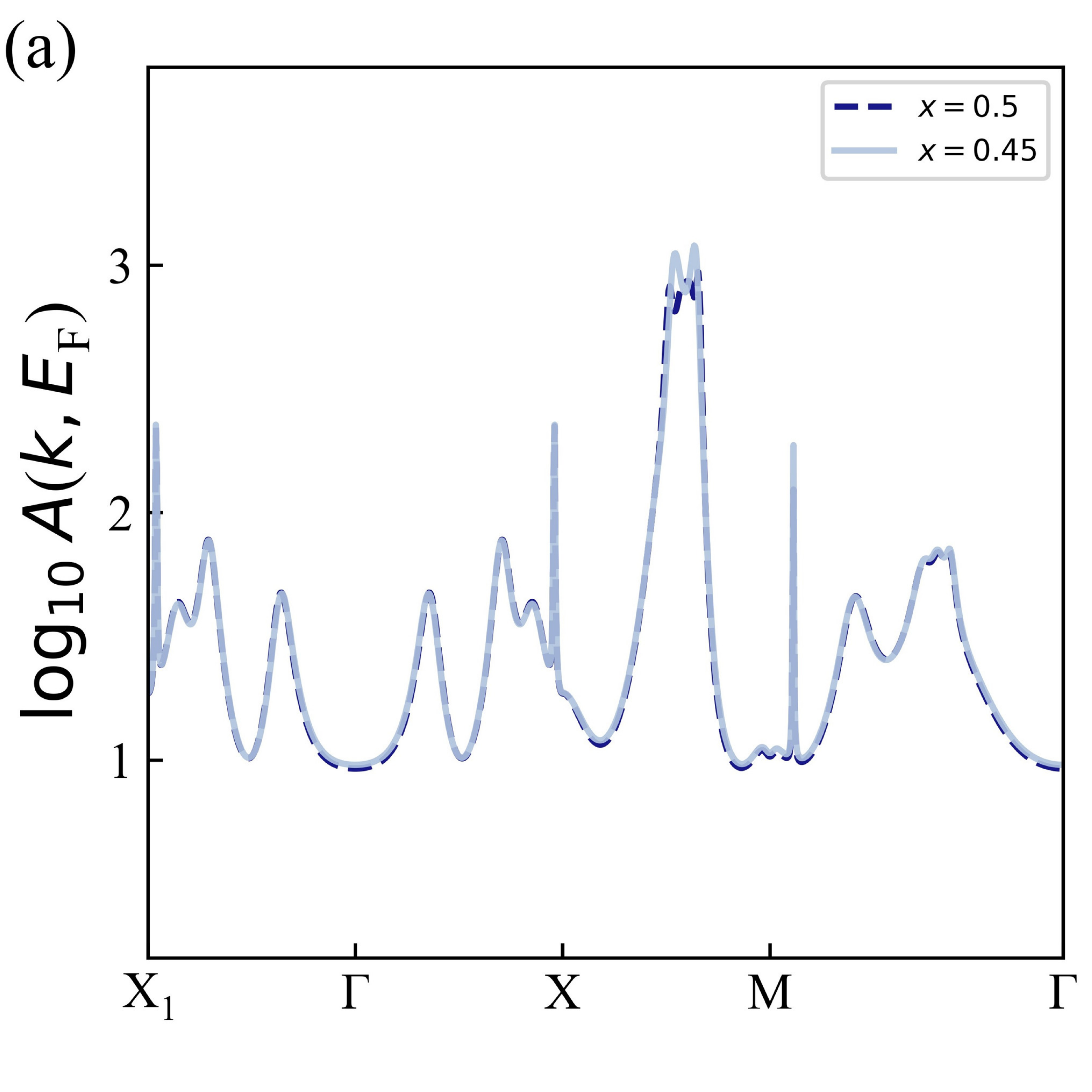}
	\includegraphics[width=0.4\linewidth]{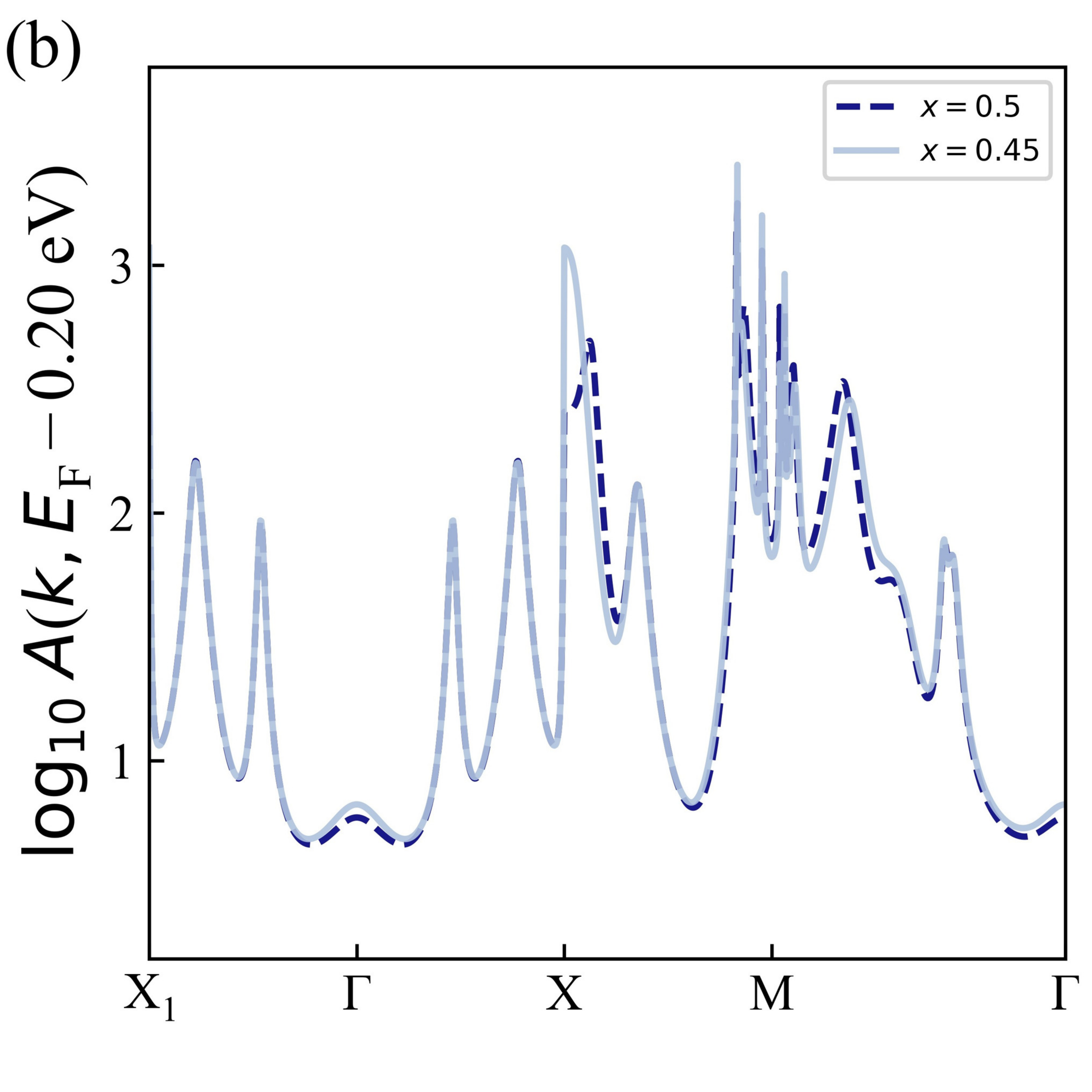}
	\caption{The BSFs of Fe$_x$(Co)$_{1-x}$Co$_x$(Fe)$_{1-x}$ with $x=0.5$ and $0.45$ with the energy levels at $E_{\mathrm{F}}$ (a) and 0.20 eV below that (b).}
	\label{fig7}
\end{figure}

In Sec. III. B, we have pointed out that the maximum values of resistivity and AMR under different anti-site disorders occur at Fe$_{0.45}$(Co)$_{0.55}$Co$_{0.45}$(Fe)$_{0.55}$ instead of Fe$_{0.5}$(Co)$_{0.5}$Co$_{0.5}$(Fe)$_{0.5}$. To find out the reason for this phenomenon, we analyzed the BSFs of the two Fe$_x$(Co)$_{1-x}$Co$_x$(Fe)$_{1-x}$ alloys at different energy in FIG. \ref{fig7} and found that they overlap in most cases. However, at -0.20 eV, the BSFs of the two diverge distinctly at the $\mathrm{X}$ and $\mathrm{M}$ points. It can be inferred that the increment of the resistivity when $x=0.45$ originates from the states 0.20 eV below the Fermi level.

\section{Conclusions}

In summary, we have investigated the influences of anti-site disorder in FeCo alloy on the AMR effect. Our work confirmed that the band-crossing, which directly determines the variations of the resistivity with respect to the magnetization direction, gives rise to the AMR. The anti-site disorder significantly enhances the scattering of the electrons, consequently increasing the resistivity, while the difference between the resistivity with magnetic moment parallel and perpendicular to the current only decreases slightly. As a consequence, the anti-site disorder affects AMR mainly by reducing conductivity. This explains the fact that suitable annealing generally enhances the AMR. The result is explained by examining the variations of the Bloch spectral function with different degrees of disorder and magnetic moment directions. 

Moreover, we fit the calculated angular dependent resistivity to the phenomenological model. We find that an alloy with a lower anti-site disorder has $\rho_{x x}$ deviating further from the $\cos 2 \alpha$ relationship. The $\cos 4 \alpha$ term rises remarkably with decreasing disorder. According to the analysis of the Bloch spectral function, the variation of the $\cos 4 \alpha$ term can be attributed to the smearing of the states near the anticrossing points in the Bloch spectral function.

\end{document}